\documentclass[preprint,epsfig]{revtex4}

\usepackage{graphicx}
\usepackage{amssymb}
\usepackage{amsmath}
\usepackage{color}
\usepackage{bm}
\usepackage{physics}
\usepackage{changes}

\newcommand{\figsizeone}{0.7}
\newcommand{\figsizetwo}{0.9}

\newcommand{\figsizefour}{0.48}

\begin{document}

\title{Optimization of conformal whispering gallery modes in lima\c{c}on-shaped transformation cavities}

\author{Jung-Wan Ryu}
\affiliation{Center for Theoretical Physics of Complex Systems, Institute for Basic Science, Daejeon 34126, Republic of Korea}
\author{Jinhang Cho}
\affiliation{Digital Technology Research Center, Kyungpook National University, Daegu 41566, Republic of Korea}
\author{Inbo Kim}
\affiliation{Digital Technology Research Center, Kyungpook National University, Daegu 41566, Republic of Korea}
\author{Muhan Choi}
\email{mhchoi@ee.knu.ac.kr}
\affiliation{Digital Technology Research Center, Kyungpook National University, Daegu 41566, Republic of Korea}
\affiliation{School of Electronics Engineering, Kyungpook National University, Daegu 41566, Republic of Korea}

\begin{abstract}

In lima\c{c}on-shaped gradient index dielectric cavities designed by conformal transformation optics, the variation of Q-factors and emission directionality of resonant modes was traced in their system parameter space. For these cavities, their boundary shapes and refractive index profiles are determined in each case by a chosen conformal mapping which is taken as a coordinate transformation. Through the numerical exploration, we found that bidirectionality factors of generic high-Q resonant modes are not directly proportional to their Q-factors. The optimal system parameters for the coexistence of strong bidirectionality and a high Q-factor was obtained for anisotropic whispering gallery modes supported by total internal reflection.
\end{abstract}

\maketitle
\narrowtext

\section{Introduction}

Whispering gallery modes (WGMs) are high-Q resonant modes supported in spherical and circular dielectric cavities, where corresponding light rays are trapped inside the cavities because the incident angle of the light rays circulating along the curved boundary are larger than the critical angle for total internal reflection (TIR) \cite{McC92, Yam93}.
However, the rotational symmetry causes isotropic light emission, which is a considerable disadvantage for applications in optical communication and integrated photonic circuits \cite{Cha96}.
Various methods have been proposed to obtain directional light emissions by breaking rotational symmetry while minimizing the spoiling of high Q-factors, for example, deformed microcavities \cite{Noe97, Gma98, Wie08, Cao15}, annular microcavities \cite{Wie06, Pre13}, coupled microcavities \cite{Ryu09, Ryu11}, and microcavities with defects at their boundaries \cite{Wan10}.

Recently, deformed gradient index microcavities designed by transformation optics \cite{Leo06, Pen06}, which are named {\it transformation cavity}, have been proposed to obtain directional light emission while simultaneously maintaining the nature of high-Q WGMs \cite{Kim16}. The cavity boundary shapes and corresponding refractive index profiles of the transformation cavities were designed utilizing conformal transformation optics \cite{Xu15}. Transformation cavities have attracted considerable attention not only in resonator optics as they combine optical microcavities with transformation optics, but also in applications requiring high-Q modes with unidirectional light emission.
There have been, however, few systematic studies on resonant mode properties in their system parameter space. Numerical investigation on the change of optical mode properties in a system parameter space is important from a practical point of view for obtaining an optimal design because the mode characteristics change substantially as the boundary shapes and corresponding refractive index profiles of cavities vary.
In this work, we studied the optical properties of resonant modes in lima\c{c}on-shaped transformation cavities. Variations in Q-factors, near-field intensity patterns, and the directionality of the far-fields of resonant modes in lima\c{c}on-shaped transformation cavities were numerically investigated as functions of system parameters.

This paper is organized as follows. We introduce a lima\c{c}on-shaped transformation cavity with an inhomogeneous refractive index profile. And then the variation in high-Q and low-Q resonant modes according to their system parameters is traced by numerical calculation. Based on these results, the optimal condition of system parameters for the resonant mode having both high Q-factor and strong bidirectional emission are obtained. Finally, we summarize our results.

\section{Results}
\subsection{Lima\c{c}on-shaped transformation cavity}

If we consider an infinite cylindrical dielectric cavity with the translational symmetry along the $z$-axis, the Maxwell equations are reduced to 2-dimensional scalar wave equation. In this case, one can use effective 2-dimensional dielectric cavity model, where optical modes are described by resonances or quasibound modes which are obtained by solving the following scalar wave equation,
\begin{equation}
[\nabla^2 + n^2 (\mathbf{r}) k^2] \psi (\mathbf{r}) = 0,
\end{equation}
where $n(\mathbf{r})$ is the refractive index function and $\mathbf{r}=(x, y)$. The resonant modes should satisfy the outgoing-wave boundary condition,
\begin{equation}
\psi(r) \sim h(\phi, k) \frac{e^{i k r}}{\sqrt{r}} ~~~~~~~~ \mathrm{for} ~~~ r \rightarrow \infty,
\end{equation}
where $h(\phi, k)$ is the far-field angular distribution of the emission.
In a conventional deformed cavity with a homogeneous refractive index profile, $n(\mathbf{r})$ is $n_0$ inside the cavity and $1$ outside the cavity. In the case of transverse magnetic (TM) polarization, the wave function $\psi(\mathbf{r})$ corresponds to $E_z$, the $z$ component of electric field \cite{Jac62}.  In the case of transverse electric (TE) polarization, the wave function $\psi(\mathbf{r})$ corresponds to $H_z$, the $z$ component of the magnetic field. Both the wave function $\psi(\mathbf{r})$ and its normal derivative $\partial_\nu \psi$ are continuous across the cavity boundary in the case of TM polarization. In the case of TE polarization, the wave function $\psi(\mathbf{r})$ is continuous across the cavity boundary and instead of its normal derivative, $n(\mathbf{r})^{-2} \partial_\nu \psi$ is continuous across the cavity boundary. The real part of the complex wave number $k$ is equal to $\omega / c$ where $\omega$ is frequency of the resonant mode and $c$ is speed of light. The imaginary part of $k$ is equal to $- 1/(2 c \tau)$ where $\tau$ is lifetime of the mode. The quality factor $Q$ of the mode is defined as $Q = 2 \pi \tau / T = - \mathrm{Re}(k) / 2 \mathrm{Im}(k)$.

\begin{figure}
\begin{center}
\includegraphics[width=\figsizeone\textwidth]{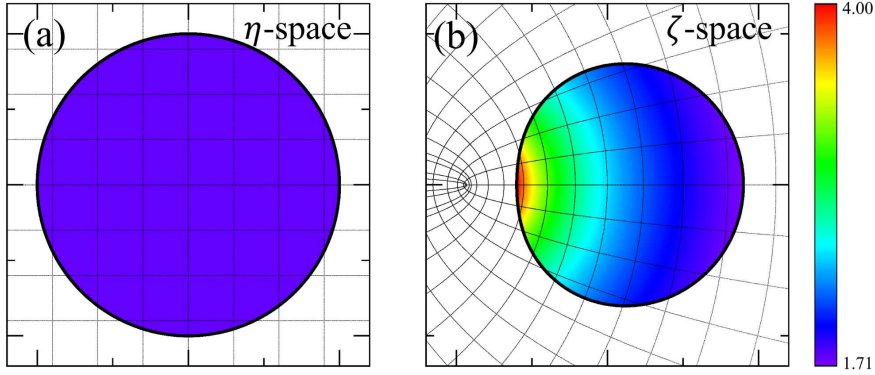}
\caption{(color online) (a) Circular dielectric cavity with homogeneous refractive index, $n_0 = 1.8$ in $\eta$-space (original virtual space). Straight gray lines denote grids of Cartesian coordinates. (b) Lima\c{c}on-shaped transformation cavity in $\zeta$-space (physical space) with inhomogeneous refractive index, which is obtained by the conformal mapping given by Eq.~(\ref{conformal}) with $\epsilon = 0.2$ and $\beta = 0.75$. Note that curved gray lines are not grids of coordinates but the transformed image of the straight grid lines in $\eta$-space by the conformal mapping. The transformed curved grids are encoded in the spatially varying refractive index inside the cavity which is denoted by scaled colors.
}
\label{fig0}
\end{center}
\end{figure}

As an example of transformation cavities, we consider a transformation cavity whose boundary is given by a lima\c{c}on shape, which is one of the widely studied shapes in the fields of quantum billiard \cite{Rob83} and optical microcavity \cite{Wie08}. The corresponding conformal mapping from the unit circle in Fig.~{\ref{fig0}} (a) to the lima\c{c}on in Fig.~{\ref{fig0}} (b) is given by
\begin{equation}
\zeta = \beta (\eta + \epsilon \eta^2),
\label{conformal}
\end{equation}
where $\eta = u + i v$ and $\zeta = x + i y$ are complex variables denoting positions in the original virtual space (see Fig.~\ref{fig0} (a)) and the physical space (see Fig.~\ref{fig0} (b)), respectively; $\epsilon$ is a deformation parameter and $\beta$ is a positive size-scaling parameter, which change the cavity boundary shape and refractive index profile of the cavity, $n(x,y) = n_0 \left| d \zeta / d \eta \right|^{-1} = n_0/\left(\beta\left|\sqrt{1+4 \epsilon \zeta / \beta}\right|\right)$, where $n_0$ is the refractive index of the unit disk cavity in the original virtual space. The refractive index outside cavity is set equal to $1$. In this work, we focus on TM polarization modes without loss of generality since TE polarization mode can be treated similarly. In the following sections, we numerically investigate the variation of resonant modes as a function of $\epsilon$ and $\beta$ using the boundary element method \cite{Wie03, Ryu17}.

\subsection{Variations of resonances according to the system parameters: high-Q modes}

First, we consider the high-Q WGM in a homogeneous circular dielectric cavity with $n_0=1.8$, whose mode number $(m, l)$ is $(14,1)$, where $m$ and $l$ are azimuthal and radial mode numbers, respectively, and the Q-factor is about $2174$. The variation of optical mode properties, including Q-factor, emission directionality, and Husimi functions as functions of the system parameters $\epsilon$ and $\beta$ are obtained. To measure the degree of bidirectional emission which is the ratio of the intensity emitted into the windows centered at $\phi=\pm \pi/2$ with an angular width of $\pi/2$ to the total emitted intensity (see the inset in Fig.~\ref{fig1} (d)), we define the bidirectionality factor B as
\begin{equation}
\mathrm{B}=\frac{\int_{\frac{\pi}{4}}^{\frac{3\pi}{4}} {I(\phi) d\phi}}{\int_{0}^{\pi} {I(\phi) d\phi}},
\label{bfactor}
\end{equation}
where $I(\phi)$ is the far-field intensity at the angle $\phi$ (see the inset in Fig.~\ref{fig1} (d)) since the distribution has the mirror symmetry with respect to the horizontal axis ($x$-axis) \cite{Ryu11}. If a mode exhibits complete isotropic emission, the B-factor is equal to $0.5$. B-factor greater than $0.5$ implies bidirectional emission in the vertical directions. When B-factor is less than $0.5$, the mode exhibits unidirectional or bidirectional emission in the horizontal directions. It should be noted that the directionality factors can be defined in other ways \cite{Son10, Shu13} or by the angle variance of far-field intensity, which yield qualitatively similar results.

\subsubsection{Q-factor and B-factor as functions of $\epsilon$ and $\beta$}

\begin{figure}
\begin{center}
\includegraphics[width=\figsizeone\textwidth]{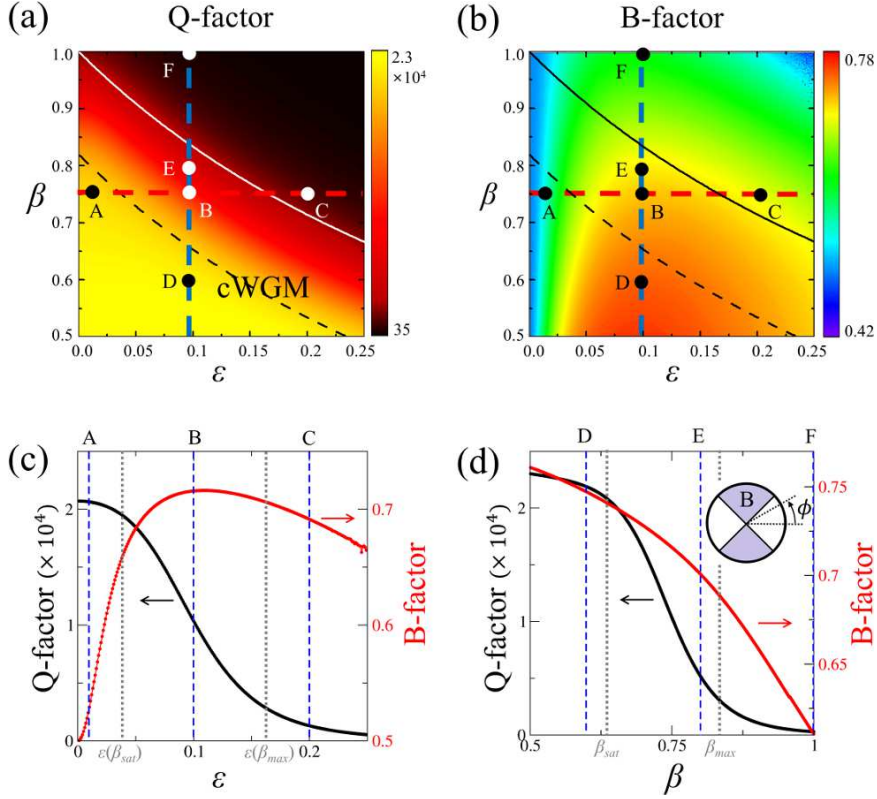}
\caption{(color online) (a) Q-factor and (b) B-factor as functions of $\epsilon$ and $\beta$.
Solid and dashed curves represent $\beta_{max}$ and $\beta_{sat}$ as a function of $\epsilon$, respectively. 
Horizontal (red  dashed) and vertical (blue  dashed) lines represent $\beta = 0.75$ and $\epsilon = 0.1$, respectively. Dots represent six selected resonant modes (A-F).
(c) Q-factor (black curve) and B-factor (red curve) as functions of $\epsilon$ with $\beta = 0.75$, corresponding to the red horizontal dashed lines in (a) and (b). Three blue dashed lines denote $\epsilon = 0.01$, $0.1$, and $0.2$.
(d) Q-factor (black curve) and B-factor (red curve) as functions of $\beta$ with $\epsilon = 0.1$, corresponding to the blue vertical dashed lines in (a) and (b). Three blue dashed lines denote $\beta = 0.6$, $0.8$, and $1.0$. The inset represents far-field angle $\phi$ and the B-factor where the shaded region indicates the range of bidirectional emission.
}
\label{fig1}
\end{center}
\end{figure}

Figure~\ref{fig1} shows Q-factor and B-factor as functions of deformation parameter $\epsilon$ and size-scaling parameter $\beta$ in lima\c{c}on-shaped transformation cavities. As deformation parameter $\epsilon$ becomes larger under a fixed $\beta$, the Q-factor typically degrades, however, the degree of degradation is much smaller as compared with severe Q-spoiling in the corresponding homogeneous cavity. On the other hand, as size-scaling parameter $\beta$ decreases when $\epsilon$ is fixed, Q-factor increases as shown in Fig.~\ref{fig1} (a) and (d), since the confinement effect becomes stronger by the enhancement of TIR mechanism. Q-factor of a mode is closely related to $\beta_{max} = 1/(1+2 \epsilon)$, the largest value of $\beta$ that support the so called, {\it conformal WGMs} (cWGMs); $\beta_{max}$ can be obtained from the condition $|d{\zeta} / d{\eta} |^{-1}  \geq 1$ necessary for TIR in transformation cavities with outside refractive index $n_{out} = 1$  \cite{Kim16}. In cases where $\beta \leq \beta_{max}$, the cWGMs with high-Q factor can be supported in transformation cavities. When $\beta > \beta_{max}$, only relatively short-lived resonances can be formed since the cWGMs are no longer supported. In Fig.~\ref{fig1} (b), for any given $\beta$ value, one can notice that B-factor is maximized at around $\epsilon = 0.1$. On the other hand, both Q-factor and B-factor have a higher value as $\beta$ value becomes smaller as shown in Fig.~\ref{fig1} (a), (b), and (d). Also, as one can see in the Q-factor curve shown in Fig.~\ref{fig1} (d), when $\beta$ is smaller than $\beta_{sat}$, the slope of the curve becomes significantly reduced. The $\beta_{sat}$ is related to the ratio of wavelength of the mode to the characteristic length scales of the system and is different for each individual mode.

\subsubsection{Change of a resonance according to $\epsilon$ variation}

\begin{figure}
\begin{center}
\includegraphics[width=\figsizetwo\textwidth]{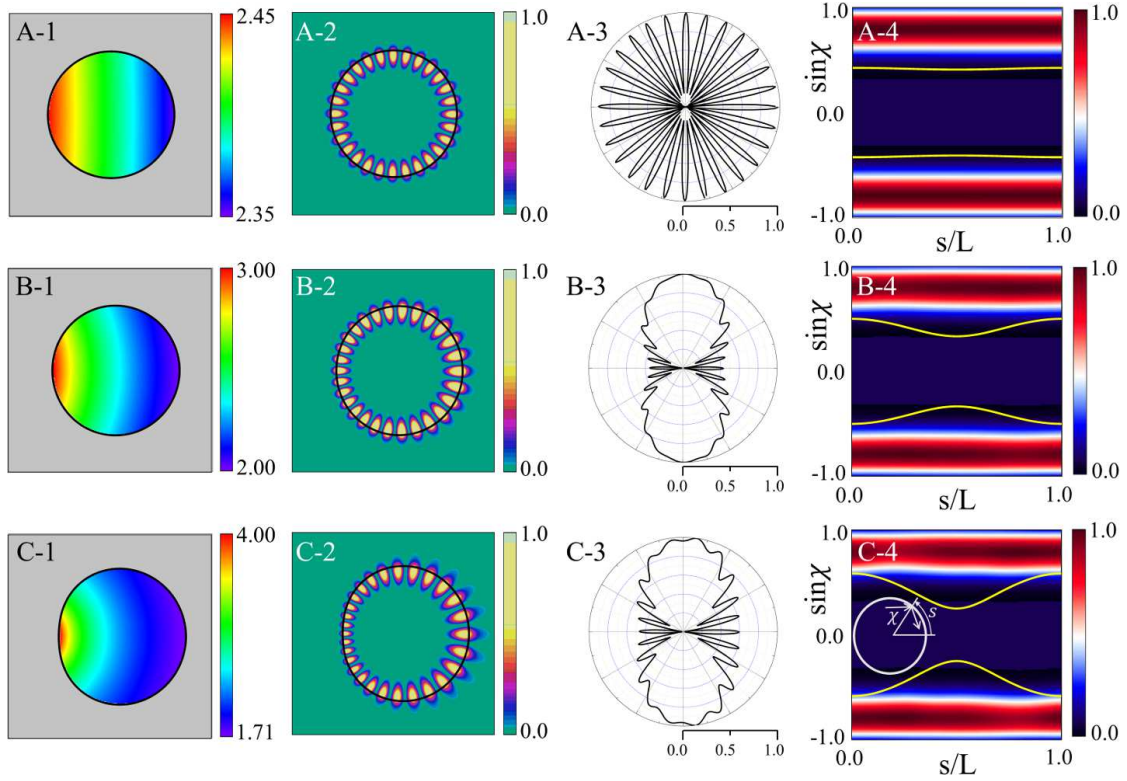}
\caption{(color online) In the first column are refractive index profiles in lima\c{c}on-shaped transformation cavities when $\epsilon$ is equal to (A-1) $0.01$ (mode-A in Fig.~\ref{fig1}), (B-1) $0.1$ (mode-B), and (C-1) $0.2$ (mode-C), respectively.
In the second column are near-field intensity patterns when $\epsilon$ is equal to (A-2) $0.01$, (B-2) $0.1$, and (C-2) $0.2$ with $\beta=0.75$, respectively.
In the third column are far-field intensity patterns when $\epsilon$ is equal to (A-3) $0.01$, (B-3) $0.1$, and (C-3) $0.2$, respectively.
In the fourth column are Husimi functions when $\epsilon$ is equal to (A-4) $0.01$, (B-4) $0.1$, and (C-4) $0.2$, respectively.
The yellow curves represent the critical lines for total internal reflection. The inset in the last Husimi function shows the arc length $s$ and incident angle $\chi$ of the ray trajectory in a reciprocal virtual space \cite{Kim18}.
}
\label{fig2}
\end{center}
\end{figure}

We investigate the variations of resonant modes as a function of $\epsilon$ for fixed $\beta = 0.75$. As can be seen in Fig.~\ref{fig1} (c), the Q-factor curve shows distinctive slope changes at two $\epsilon$ values associated with $\beta_{sat}$ and $\beta_{max}$. The reason of the slope change at $\epsilon \sim 0.167$ is that $\beta$ starts to violate the TIR condition ($\beta > \beta_{max}(\epsilon) = 0.75$) when $\epsilon > 0.167$. We will discuss similar behavior in the next section on the $\beta$ parameter variation. The B-factor has a maximum value around $\epsilon = 0.1$. The reason for the change of emission directionality according to the deformation can be easily understood by plotting the Husimi function \cite{Hen03, Lee05}, one of the widely used phase space representation of intracavity wave intensity. For our transformation cavities, the Husimi function can be calculated in the reciprocal virtual space which is obtained by an inverse conformal mapping from the physical space \cite{Kim18}. Figure~\ref{fig2} shows refractive index profiles, near-field intensity patterns, far-field intensity patterns, and Husimi functions for the modes with $\epsilon = 0.01$, $0.1$, and $0.2$ (A, B, and C marked in Fig.~\ref{fig1}, respectively). In the case with $\epsilon = 0.01$ (very small deformation), the range of the refractive index profile is narrow as shown in Fig.~\ref{fig2} (A-1), so the near and the far-field intensity patterns of the mode-A depicted in Fig.~\ref{fig2} (A-2) and (A-3) are very similar as those of the corresponding WGM with isotropic emission in a uniform circular cavity. Also, in the Husimi function depicted in Fig~\ref{fig2} (A-4), the upper and lower bands above the critical line which represent the intensities of counterclockwise (CCW) and clockwise (CW) traveling wave components, respectively, is slightly changed from the straight uniform bands of the Husimi function of the WGM in a uniform circular cavity. The distance between the intensity tails of bands of Husimi function and critical line is nearly the same at all positions of cavity boundary. Thus the mechanism of the almost isotropic emission is direct tunneling.

When $\epsilon$ becomes $0.1$, the far-field intensity pattern of the mode-B exhibits a pronounced bidirectional emission as shown in Fig.~\ref{fig2} (B-3) while maintaining the Q-factor sufficiently high as shown in Fig.~\ref{fig1}. The band-type intensities of Husimi function becomes distorted a little bit more because of the breaking of rotational symmetry, but are still very similar to those of a WGM in a uniform circular cavity, unlike the cases of conventional deformed cavities with homogeneous refractive index. On the other hand, the critical lines for TIR become bent from straight lines into curved ones as shown in Fig~\ref{fig2} (B-4). The dominant emissions leak out in two opposite tangential directions slightly off the cavity boundary position ($s=0$) where the intensity tails of the CW and CCW wave bands of Husimi functions are closest to the critical lines (i.e., where the refractive index ratio between inside and outside cavity is lowest). Since the distance between the intensity bands of Husimi function and critical lines is sufficiently away, emission mechanism is still direct tunneling. The main reason that the far-field intensity patterns drastically change from almost isotropic (mode-A) to bidirectional emission patterns (mode-B) even though the near-field intensity pattern of each cWGM is similar to those of WGM in uniform index circular cavities is the bending down and up of the critical lines, i.e., symmetric rise and fall of in/out index ratio along the cavity boundary, not the variations of the band structure of Husimi function of cWGMs. When $\epsilon = 0.2$, the mode-C is not a cWGM since $\beta$ is larger than $\beta_{max}$ as shown in Fig.~\ref{fig1} and then the Q-factor becomes lower since the distance between the intensity bands of Husimi function and critical lines is closer at $s=0$ than the previous case, as can be noticed in Fig.~\ref{fig2} (C-4). In spite of lower Q-factor, the far-field intensity pattern of the mode-C still exhibits a bidirectional feature as shown in Fig.~\ref{fig2} (C-3). To summarize, in a lima\c{c}on-shaped transformation cavity, as the deformation parameter $\epsilon$ starts to increase for a fixed $\beta$ value, the high-Q WGM with isotropic emission in a circular dielectric cavity changes into the high-Q cWGM with bidirectional emission. As $\epsilon$ increases further to break the TIR condition, the high-Q cWGM transforms into the low-Q mode with some degraded bidirectionality.

\subsubsection{Change of a resonance according to $\beta$ variation}

\begin{figure}
\begin{center}
\includegraphics[width=\figsizetwo\textwidth]{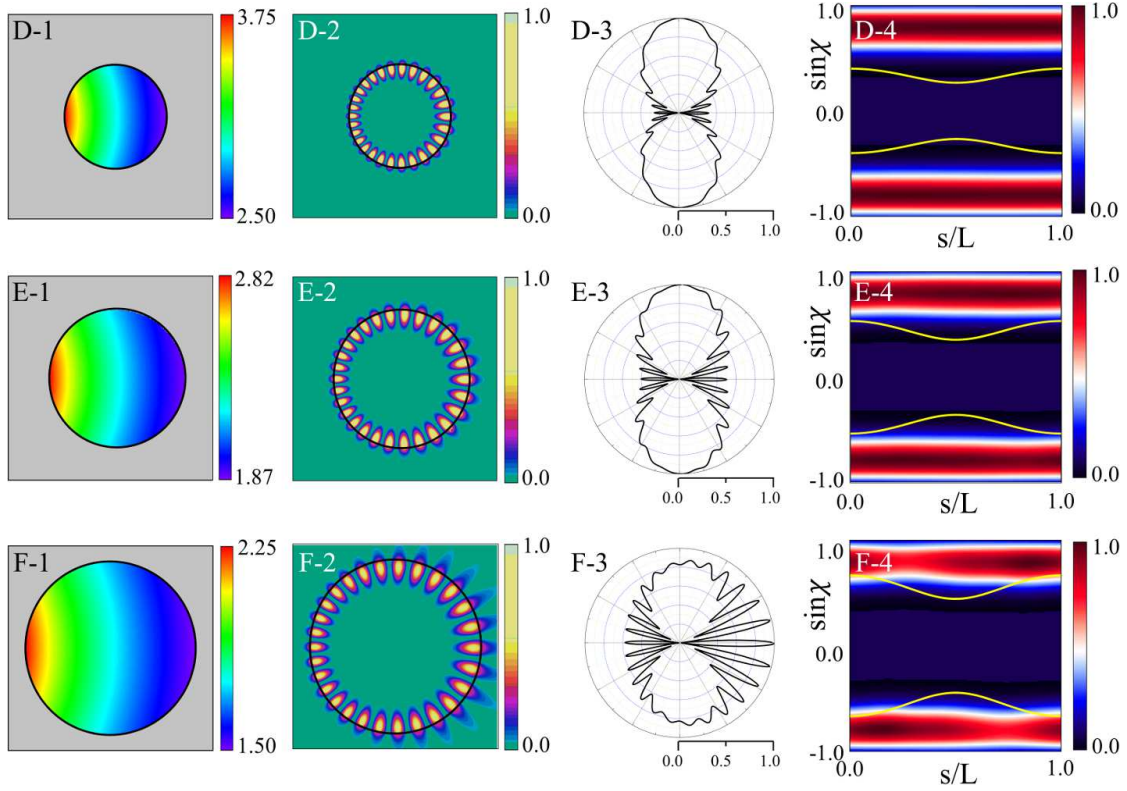}
\caption{(color online) 
In the first column are refractive index profiles in lima\c{c}on-shaped transformation cavities when $\beta$ is equal to (D-1) $0.6$ (mode-D in Fig.~\ref{fig1}), (E-1) $0.8$ (mode-E), and (F-1) $1.0$ (mode-F), respectively.
In the second column are near-field intensity patterns when $\beta$ is equal to (D-2) $0.6$, (E-2) $0.8$, and (F-2) $1.0$ with $\epsilon=0.1$, respectively.
In the third column are far-field intensity patterns when $\beta$ is equal to (D-3) $0.6$, (E-3) $0.8$, and (F-3) $1.0$, respectively.
In the fourth column are Husimi functions when $\beta$ is equal to (D-4) $0.6$, (E-4) $0.8$, and (F-4) $1.0$, respectively.
}
\label{fig3}
\end{center}
\end{figure}

Next, we investigate the variations of resonant modes as a function of $\beta$ for fixed $\epsilon = 0.1$. As shown in Fig.~\ref{fig1} (d), as $\beta$ decreases, Q-factor of the mode increases monotonically through two characteristic $\beta$ values where the slope of Q-factor curve is significantly changed. Approaching to the first point, $\beta_{max} = 0.833$, Q-factor of the mode steeply reaches a very high value because it enters to the parametric domain satisfying the TIR condition, $\beta \leq \beta_{max} ~\sim 0.833$. This tendency is also seen in Fig.~\ref{fig1} (c) where the Q-factor begins to increase from $\epsilon \sim 0.167$ when $\beta = 0.75$. Reaching around the second point $\beta_{sat} \sim 0.63$, the Q-factor of the mode starts to be saturated because most of the plane wave components of the incident wave faithfully undergoes TIR. This means that in order to keep the Q-factor value of the mode high, $\beta$ should be set sufficiently smaller than $\beta_{max}$.

Figure~\ref{fig3} shows refractive index profiles, near-field intensity patterns, far-field intensity patterns, and Husimi functions for the modes with $\beta = 0.6$, $0.8$, and $1.0$ (D, E, and F marked in Fig.~\ref{fig1}), respectively. The mode-D and E with $\beta < \beta_{max}$ are cWGMs and their far-field intensity patterns exhibit bidirectional emission as shown in Figs.~\ref{fig3} (D-3) and (E-3). The mode-F with $\beta > \beta_{max}$ is not a cWGM and its bidirectionality of emission is considerably spoiled as shown in Fig.~\ref{fig3} (F-3). The B-factor monotonically decrease as $\beta$ increase in Fig.~\ref{fig1} (d) unlike the case of $\epsilon$ variation shown in Fig.~\ref{fig1} (c). As shown in Fig.~\ref{fig3} (D-4), when $\beta = 0.6$, the upper and lower bands of the Husimi function of the mode-D is far from the critical lines. When $\beta = 0.8$, the lines become closer to the bands as shown in Fig.~\ref{fig3} (E-4). When $\beta = 1.0$ violating the TIR condition, the critical lines for the mode-F overlap with the bands over a fairly large region centered around $s=0$, as depicted in Fig.~\ref{fig3} (F-4).

Through the above investigation for the change of mode characteristics, we can know that, as $\beta$ crosses $\beta_{max}$ of TIR condition, the high-Q WGM can change between a low-Q mode and a cWGM. Also a higher Q-factor of the cWGM accompanies better bidirectional emission as $\beta$ varies when $\epsilon$ is fixed. Additionally, smaller $\beta$ requires the transformation cavity have wider range of refractive index profile with a higher maximum index value. The range of refractive index profile is an important restriction in actual fabrication of the transformation cavity, so the optimal $\beta$ should be selected appropriately within the attainable range of refractive index.

\begin{figure}
\begin{center}
\includegraphics[width=\figsizefour\textwidth]{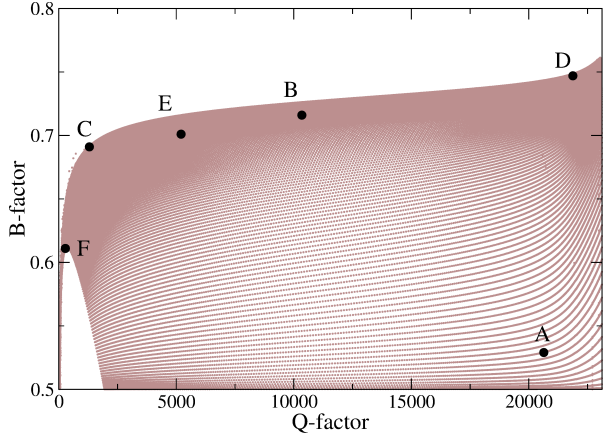}
\caption{(color online) Diagram of Q-factor vs. B-factor space of resonant modes (brown dots) in the regions of $0 \leq \epsilon < 0.25$ and $0.5 < \beta < 1.0$. The large black dots represent six resonant modes (A-F).
}
\label{fig3-2}
\end{center}
\end{figure}

Figure~\ref{fig3-2} shows the diagram which represent the positions of the resonant modes depicted in Fig.~\ref{fig1} in the Q-factor vs. B-factor space. In this diagram, one can see the parameters for the resonant mode-D is a nearly optimal condition for high Q-factor and strong bidirectional emission in a lima\c{c}on-shaped transformation cavity with $0 \leq \epsilon < 0.25$ and $0.5 < \beta < 1.0$. In the case of a cWGM with a sufficiently high Q-factor, the major emission occurs in the two opposite tangential directions from the one small spot slightly off the cavity boundary position ($s = 0$) where the in/out refractive index ratio is smallest. Therefore, in a lima\c{c}on-shaped transformation cavity, tunneling emission occurs only at one position ($s = 0$) of the cavity boundary for all cWGMs, i.e., the emission directionality of a high-Q cWGM is universal. However, in contrast to high-Q cWGMs, the far-field intensity patterns of low-Q modes are not universal, but instead represent specific properties of individual modes, which will be dealt with briefly in the following section.

\subsection{Variations of resonances depending on the system parameters: low-Q modes}

In this section, we consider a low-Q mode in a circular dielectric cavity with $n_0 = 1.8$, whose mode number $(m,l)$ is $(8,3)$ and the Q-factor is about $84$. Just as in the case of a high-Q cWGM, we also study the variation of the mode and its optical properties, such as Q-factor, emission directionality, and Husimi function, as functions of the system parameters in a lima\c{c}on-shaped transformation cavity. For this mode, as the deformation parameter increases, we obtain unidirectional emission, which is a specific property of an individual mode. To measure the degree of unidirectional emission which is the ratio of the intensity emitted into a $\pi/2$-degree window centered around $\phi=0$ to the total emitted intensity (see the inset in Fig.~\ref{fig4} (d)), similarly to the B-factor in the previous section, we define a unidirectionality factor U as
\begin{equation}
\mathrm{U}=\frac{\int_{0}^{\frac{\pi}{4}} {I(\phi) d\phi}}{\int_{0}^{\pi} {I(\phi) d\phi}},
\label{ufactor}
\end{equation}
where $I(\phi)$ is the far-field intensity distribution at the angle $\phi$ since the wave function has mirror symmetry for horizontal axis ($x$-axis). The U-factor is equal to $0.25$ when a mode exhibits complete isotropic emission. Thus, when the U-factor is larger than $0.25$, the mode exhibits unidirectional emission in the horizontal direction.

\subsubsection{Q-factor and U-factor as functions of $\epsilon$ and $\beta$}

\begin{figure}
\begin{center}
\includegraphics[width=\figsizeone\textwidth]{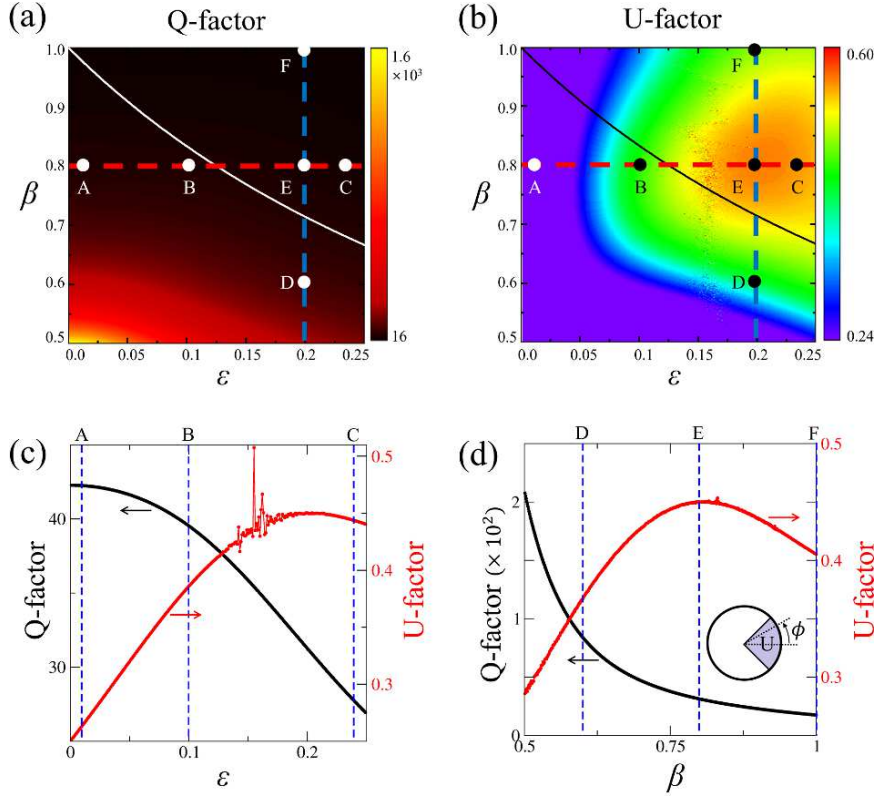}
\caption{(color online) (a) Q-factor and (b) U-factor as functions of $\epsilon$ and $\beta$. Solid curves represent $\beta_{max}$ as a function of $\epsilon$. 
Horizontal (red dashed) and vertical (blue dashed) lines represent $\beta = 0.8$ and $\epsilon = 0.2$, respectively. Dots represent six selected resonant modes (A-F).
(c) Q-factor (black curve) and U-factor (red curve) as functions of $\epsilon$ with $\beta = 0.8$, corresponding to the red horizontal dashed lines in (a) and (b). Three blue dashed lines denote $\epsilon = 0.01$, $0.1$, and $0.24$.
(d) Q-factor (black curve) and B-factor (red curve) as functions of $\beta$ with $\epsilon = 0.2$, corresponding to the blue vertical dashed lines in (a) and (b). Three blue dashed lines denote $\beta = 0.6$, $0.8$, and $1.0$. The shaded region of the inset indicates the range of unidirectional emission.
}
\label{fig4}
\end{center}
\end{figure}

Figure~\ref{fig4} shows the Q-factor and U-factor as functions of $\epsilon$ and $\beta$. As $\beta$ decreases, the Q-factor, of course, increases due to the overall rising of refractive index as mentioned above. In contrast to a high-Q cWGM, the Q-factor of the mode is not strongly related to $\beta_{max}$ because low-Q modes do not satisfy the TIR condition. The aspect of U-factor variation as functions of $\epsilon$ and $\beta$ differs from that of Q-factor variation. As $\beta$ increases, the U-factor of the mode increases in the region of $\beta < 0.8$, but decreases in the region of $\beta > 0.8$. This means that there is critical value of $\beta$ for unidirectionality of emission as shown in Fig.~\ref{fig4} (d). A highest U-factor is obtained near $\epsilon = 0.2$ and $\beta = 0.8$ as shown in Fig.~\ref{fig4}.

\subsubsection{Change of a resonance according to $\epsilon$ variation}

\begin{figure}
\begin{center}
\includegraphics[width=\figsizeone\textwidth]{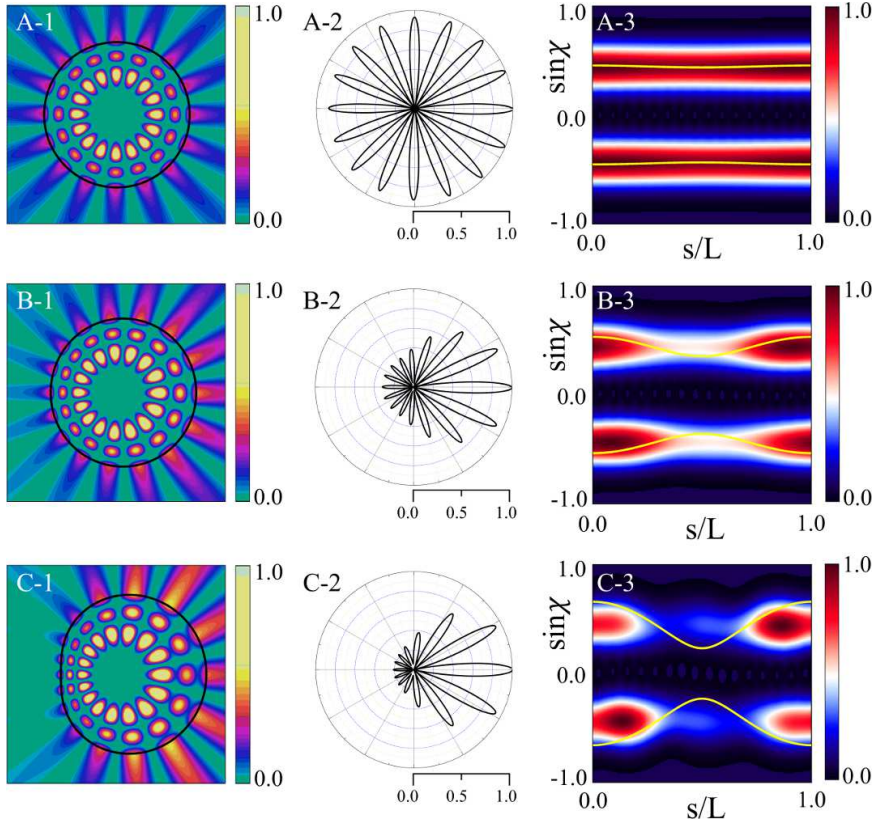}
\caption{(color online) 
In the first column are near-field intensity patterns in a lima\c{c}on-shaped transformation cavity when $\epsilon$ is equal to (A-1) $0.01$ (mode-A in Fig.~\ref{fig4}), (B-1) $0.1$ (mode-B), and (C-1) $0.24$ (mode-C) when $\beta=0.8$, respectively.
In the second column are far-field intensity patterns when $\epsilon$ is equal to (A-2) $0.01$, (B-2) $0.1$, and (C-2) $0.24$, respectively.
In the third column are Husimi functions when $\epsilon$ is equal to (A-3) $0.01$, (B-3) $0.1$, and (C-3) $0.24$, respectively.
Yellow curves represent the critical lines for total internal reflection. 
}
\label{fig5}
\end{center}
\end{figure}

We now investigate the variations of resonant modes as a function of $\epsilon$ when $\beta=0.8$. As $\epsilon$ increases, the Q-factor decreases and the U-factor increases for $\epsilon \lesssim 0.2$, but decreases gradually for $\epsilon \gtrsim 0.2$. Figure~\ref{fig5} shows the near-field intensity patterns and far-field intensity patterns as well as the Husimi functions when $\epsilon=0.01$, $0.1$, and $0.24$. The refractive index profiles are similar to those in Fig.~\ref{fig2} and Fig.~\ref{fig3}. When $\epsilon = 0.01$, the near-field intensity patterns are nearly the same as those of a low-Q mode with a mode number of $(m, l) = (8,3)$ in a homogeneous circular cavity, and the far-field intensity patterns are nearly isotropic. When $\epsilon = 0.1$ and $0.2$, the U-factor is large because of unidirectional emission, which is a specific property of an individual mode, in contrast to the universal bidirectional emission of a high-Q cWGM. The far-field intensity pattern shows clear unidirectional emission, as shown in Figs.~\ref{fig5} (B-2) and (C-2). The Husimi functions in Figs.~\ref{fig5} (A-3), (B-3), and (C-3) show that the Q-factor of the modes becomes lower as the CW/CCW wave intensities of Husimi functions move below the critical lines for the TIR. It should be noted that there are regions with fluctuating U-factors in Fig.~\ref{fig4} (b) and (c) when $0.14 \lesssim \epsilon \lesssim 0.18$ where the Q-factor changes smoothly but the far-field intensity pattern changes rapidly.

\subsubsection{Change of a resonance according to $\beta$ variation}

\begin{figure}
\begin{center}
\includegraphics[width=\figsizeone\textwidth]{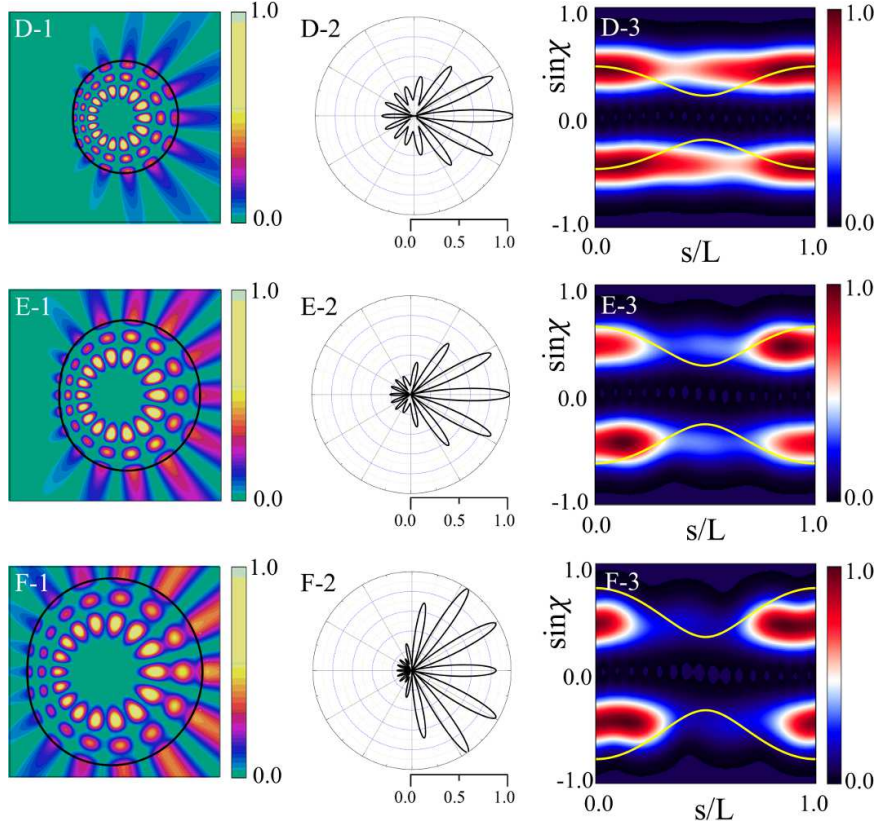}
\caption{(color online)
In the first column are near-field intensity patterns in a lima\c{c}on-shaped transformation cavity when $\beta$ is equal to (D-1) $0.6$ (mode-D in Fig.~\ref{fig4}), (E-1) $0.8$ (mode-E), and (F-1) $1.0$ (mode-F) when $\epsilon=0.2$, respectively.
In the second column are far-field intensity patterns when $\beta$ is equal to (D-2) $0.6$, (E-2) $0.8$, and (F-2) $1.0$, respectively.
In the third column are Husimi functions when $\beta$ is equal to (D-3) $0.6$, (E-3) $0.8$, and (F-3) $1.0$, respectively.
}
\label{fig6}
\end{center}
\end{figure}

We investigate the variations of resonant modes as a function of $\beta$ when $\epsilon=0.2$. As $\beta$ increases, the Q-factor decreases, but the U-factor increases when $\beta$ is less than $0.8$ and decreases when $\beta$ is greater than $0.8$. Figure~\ref{fig6} shows the near-field intensity patterns and far-field intensity patterns as well as the Husimi functions when $\beta=0.6$, $0.8$, and $1.0$, for $\epsilon=0.2$. They are similar to those shown in Fig.~\ref{fig5}.

\section{Summary}

We studied the optical properties of the resonant modes in lima\c{c}on-shaped transformation cavities with an inhomogeneous refractive index profiles by changing system parameters. From numerical calculations of Q-factors, the directionality factors of the far-fields, and Husimi functions of resonant modes as functions of system parameters, we found that generic high-Q resonant modes exhibit bidirectional emissions but the bidirectionality factors of the modes are not directly proportional to their Q-factors. In contrast to the universal bidirectional emission of a high-Q cWGM, the directionality of low-Q modes are the specific property of individual mode.
In the implementation of a transformation cavity, the maximum and minimum values of the refractive index inside the cavity are limited by the lack of a high refractive index material in nature. We demonstrated that lima\c{c}on-shaped transformation cavities which support optimal cWGM with high-Q factor and strong bidirectional emission can be designed within an attainable range of refractive index. We expect that our approach and results for transformation cavities will be useful to design of advanced optical devices.

\section*{Acknowledgment}
This work was supported by the Institute for Basic Science of Korea (IBS-R024-D1) and National Research Foundation of Korea (NRF) grant funded by the Korea government (MSIP) (No. 2017R1A2B4012045 and No. 2017R1A4A1015565).

\section*{Author contributions statement}

J.-W.R., J.C., I.K., and M.C. conceived the original idea. J.-W.R., and J.C. performed numerical simulations. J.-W.R., J.C., I.K., and M.C. analyzed the data and discussed the results. All authors wrote the manuscript and provided feedback.

\section*{Competing interests}
The authors declare no competing interests.

\end{document}